\documentclass[aps,preprint]{revtex4}%
\usepackage{amsfonts}
\usepackage{amsmath}
\usepackage{amssymb}
\usepackage{graphicx}%
\providecommand{\U}[1]{\protect\rule{.1in}{.1in}}

\begin{document}
\title[ ]{Classical Zero-Point Radiation and Relativity: The Problem of Atomic Collapse Revisited}
\author{Timothy H. Boyer}
\affiliation{Department of Physics, City College of the City University of New York, New
York, New York 10031}

\begin{abstract}
The physicists of the early 20th century were unaware of two aspects which are
vital to understanding some aspects of modern physics within classical theory.
\ The two aspects are: 1) the presence of classical electromagnetic zero-point
radiation, and 2) the importance of special relativity. \ In classes in modern
physics today, the problem of atomic collapse is still mentioned in the
historical context of the early 20th century. \ However, the classical problem
of atomic collapse is currently being treated in the presence of classical
zero-point radiation where the problem has been transformed. The presence of
classical zero-point radiation indeed keeps the electron from falling into the
Coulomb potential center. \ However, the old collapse problem has been
replaced by a new problem where the zero-point radiation may give too much
energy to the electron so as to cause \textquotedblleft
self-ionization.\textquotedblright\ \ Special relativity may play a role in
understanding this modern variation on the atomic collapse problem, just as
relativity has proved crucial for a classical understanding of blackbody radiation.

\end{abstract}
\maketitle

\section{Introduction}

In 1911 when Rutherford used the data of scattering experiments to publish the
nuclear model of the atom, the problem of atomic collapse immediately arose.
Earlier experimental work, notably by J. J. Thomson, had measured the ratio of
$e/m$ for the electron. The work on the normal Zeeman effect by Zeeman and
Lorentz placed electrons as part of the atom and as connected to spectral
lines. But Rutherford's model of the atom involving electrons in Kepler-like
orbits around a small nucleus raised a profound question: the orbiting
electrons were obviously accelerating, and, according to classical
electromagnetic theory, the electrons must radiate away their energy; what
prevented the collapse of the electron into the nucleus? \ This
\textquotedblleft problem of atomic collapse\textquotedblright\ is always
mentioned in our classes in modern physics when presenting the need for a
quantum description of atomic physics.\cite{R-B} \ 

In 1913, Bohr \textquotedblleft solved\textquotedblright\ the problem of
atomic collapse by fiat. The old quantum theory claimed simply that in certain
stationary orbits, the electrons did not obey the laws of classical
electromagnetism and did not radiate. Currently, the stationary orbits of old
quantum theory have been replaced by the stationary states of Schroedinger
quantum mechanics. However, the outlook introduced by Bohr in 1913 remains; a
new quantum theory must replace the ideas of classical physics when dealing
with the structure of the atom. \ 

In contrast to the historical view presented in our modern physics courses,
today the century-old mystery of atomic collapse within classical physics is
taking a fascinating new twist. \ It turns out that there are two significant
ideas which were absent from the thinking of the early 20th century physicists
which might have helped the classical theoreticians of the period. These two
ideas are 1) the presence of classical electromagnetic zero-point radiation in
the universe, and 2) the importance of special relativity. When these two
ideas are introduced into the old problem of atomic collapse, the problem is
transformed. In recent years, there have been attempts to solve the problem of
atomic collapse when classical zero-point radiation is present.\cite{CZ}%
\cite{NL1}\cite{NL2}

Relativistic classical electron theory with classical electromagnetic
zero-point radiation (sometimes termed \textquotedblleft stochastic
electrodynamics\textquotedblright) is the classical theory which most closely
approximates quantum physics. \ Clearly, physicists would like to have a clear
picture of the areas of agreement and disagreement between classical and
quantum theories. \ Recently Nieuwenhuizen and Liska\cite{NL1}\cite{NL2} have
undertaken a heroic effort to extend the fascinating numerical calculations of
Cole and Zou\cite{CZ} regarding the ground state of hydrogen when classical
zero-point radiation is present. \ Here we wish to review these efforts as an
interesting attempt to strengthen our understanding of the classical aspects
of atomic physics.

\section{The Hydrogen Ground State in Classical Physics}

\subsection{Classical Electromagnetic Zero-Point Radiation}

Classical electrodynamics is a well-established theory which is the basis for
much of modern technology; it has a well-defined framework of differential
equations and boundary conditions. In addition to describing the charges and
currents which appear as sources for the electric and magnetic fields in
Maxwell's equations, one must also choose the homogeneous boundary conditions
on the equations. The situation seems obvious if one considers the
electromagnetic radiation in a laboratory before the experimenter's sources
are turned on. \ As far as the experimenter is concern, his sources did not
provide the radiation in the lab which existed when he entered. \ This
radiation which does not arise from the experimenter's sources can be treated
as boundary conditions appearing in the solution of the homogeneous Maxwell
equations. \ The thermal radiation present in the laboratory is an example of
random classical radiation which would be treated as a solution of the
homogeneous Maxwell equations. \ Every classical electromagnetic theory based
upon Maxwell's differential equations must make an assumption regarding the
homogeneous boundary conditions on these differential equations. \ In the
early years of the 20th century, the assumption was made that the homogeneous
boundary condition corresponded to a complete absence of radiation before the
sources acted. \ Lorentz makes this assumption explicitly in his book on
classical electron theory.\cite{L} \ A different but also legitimate classical
boundary condition on Maxwell's equations involves the presence of random
classical radiation with a Lorentz-invariant spectrum, classical
electromagnetic zero-point radiation.\cite{texts}\ \ \ Classical
electromagnetic zero-point radiation is allowed by classical electromagnetic
theory and provides a classical understanding of some aspects of nature.

The spectrum of \ classical zero-point radiation is derived as the
Lorentz-invariant spectrum of random classical electromagnetic
radiation.\cite{M1965}\cite{B-rel} \ The spectrum is unique up to a
multiplicative constant as an energy $\mathcal{E}$ per normal mode of
frequency $\omega$ given by
\begin{equation}
\mathcal{E}=const\times\omega
\end{equation}
When this Lorentz-invariant spectrum is used to calculate Casimir forces, it
is found that the scale constant for classical zero-point radiation must be
chosen as $const=1.05\times10^{-34}$J$\cdot$s in order to fit the experimental
data. \ Although the $const$ here has nothing to do with quanta, the value of
$const$ clearly corresponds to $\hbar/2$ where $\hbar$ is Planck's constant
appearing in quantum physics. \ Thus Planck's constant $\hbar$ enters
classical theory simply as the scale factor of classical zero-point radiation.
\ Today we regard classical physics with classical zero-point radiation as the
closest classical approximation to quantum physics.\cite{B-close} \ 

\subsection{Review of the Basic Idea for Hydrogen}

It is easy to calculate the behavior of a linear dipole oscillator in
classical zero-point radiation.\cite{Rev1}\cite{Rev2} \ The dipole oscillator
radiates away its energy but it also picks up energy from the random forces of
the classical zero-point radiation and so comes to a steady state probability
distribution for energy, amplitude, and velocity. \ Using classical zero-point
radiation, it turns out that Casimir forces, van der Waals forces, blackbody
radiation, low-temperature specific heats of solids, and diamagnetism all come
within the framework of this classical theory.\cite{B-close}\ 

Although harmonic oscillator systems and systems of free fields are easy to
calculate in the presence of classical zero-point radiation, the hydrogen atom
is a far more complicated system. \ In 1975, when the results of classical
electrodynamics with classical zero-point radiation were reviewed, it was
pointed out that there was a clear qualitative suggestion that the classical
theory might lead to a stable hydrogen ground state.\cite{Rev1} \ It was
reported that if one approximated the ground-state motion as analogous to that
of a nonrelativistic, planar, rigid rotator where a particle of charge $e$ and
mass $m$ is held at a fixed distance $r$ from the rotation center, then the
power loss $d\mathcal{E}_{loss}/dt$ due to radiation emission was given by%
\begin{equation}
\frac{d\mathcal{E}_{loss}}{dt}=\frac{2}{3}\frac{e^{2}}{c^{3}}\omega^{4}r^{2},
\end{equation}
whereas the average energy pick-up $d\mathcal{E}_{gain}/dt$\ from the random
classical zero-point radiation was given by
\begin{equation}
\frac{d\mathcal{E}_{gain}}{dt}=\frac{e^{2}\hbar\omega^{3}}{2mc^{3}},
\end{equation}
\ where here $\hbar$ is the scale factor setting the scale of the classical
zero-point energy. \ \ At high frequencies $\omega$, the particle lost more
energy by radiation than it picked up from the zero-point radiation; on the
contrary, at low frequencies, the oscillator picked up more energy than it
radiated away. \ Thus one expected a ground state at the balance point
corresponding
\begin{equation}
\frac{d\mathcal{E}_{loss}}{dt}=\frac{d\mathcal{E}_{gain}}{dt}\text{ \ or
\ }mr^{2}\omega=J=\frac{3}{4}\hbar
\end{equation}
where $J$ is the angular momentum of the rotator. \ Old quantum theory chose
the result $J=\hbar.$ \ Thus this qualitative model for hydrogen gives
approximately the same result as that for the Bohr-model ground state or for
the Schroedinger ground state. \ However, this rough, heuristic estimate is
far from a real calculation.\cite{calc}

Within classical physics, the basic picture for the hydrogen ground state
involves a separation between a mechanical system of a particle in a Coulomb
potential $Ze^{2}/r$ and a spectrum of random classical radiation with a
Lorentz-invariant spectrum. \ Both these systems can be described by
action-angle variables.\cite{action} The slowly-changing action-angle
variables provide the appropriate variables for a canonical perturbation
theory between the mechanical system and the random radiation.\ The mechanical
orbits of the particle in the Coulomb potential are perturbed by the loss of
energy to radiation and also by the random forces of the zero-point radiation.
\ The zero-point radiation field is increased by the energy radiated by the
charged particle and is diminished by the absorption of radiation by the
mechanical system. \ 

\section{Current Research on the Hydrogen Ground State}

\subsection{Calculations of Cole and Zou}

Although the basic physical picture for the classical hydrogen atom is clear,
there is at present no full analytic calculation of the hydrogen ground state
when zero-point radiation is present. \ Rather, the evaluation of the
electron's motion involves enormously difficult computer simulations. \ The
calculations are extremely difficult because of the need to simulate the
spectrum of random classical zero-point radiation and because of the need to
follow the perturbed electron through vast numbers of orbits. \ In 2003, Cole
and Zou reported the first computer simulation for the hydrogen ground
state.\cite{CZ} \ In order to simplify the calculation, they restricted the
mechanical particle motion to a single plane. \ Their nonrelativistic
calculations were quite favorable to the classical theory. In no case did the
electron fall into the Coulomb center. Also, the numerical calculations
suggested that the probability distribution for the radial distance of the
electron from the hydrogen nucleus roughly approximated that given by the
Schroedinger ground state of quantum theory. \ In situations where the
numerical simulations suggested that the energy gain from zero-point radiation
was sufficiently large as to ionize the atom, more accurate recalculations
showed that the ionization actually did not occur.

\subsection{Claverie and Soto Regarding Ionization}

Indeed, the question of self-ionization of the hydrogen atom in classical
zero-point radiation appears as a new aspect of classical atomic structure.
\ Whereas the traditional classical theory (without zero-point radiation) of
the early 20th century suggested atomic collapse because the accelerating
electron would radiate away its energy, the new classical theory (which
included zero-point radiation) raises the possibility that the electron might
acquire sufficient energy from the zero-point radiation so as to spontaneously
ionize. \ Work by Marshall, Claverie, Pesquera, and Soto\cite{MC} (using
nonrelativistic mechanics for the electron) suggested that there was no stable
ground state for hydrogen; the electron would always be ionized by the
zero-point radiation. \ Presumably the nonrelativistic calculations of Cole
and Zou did not continue sufficiently long so as to show this self-ionization.

The ionization found in the nonrelativistic classical theory always involved
the plunging orbits of low angular momentum. \ However, it was pointed out
that these plunging orbits of small angular momentum are precisely the orbits
which are strongly modified by using relativistic theory for the mechanical
motion of the electron.\cite{B-h} \ Thus whereas analytic calculations using
\textit{nonrelativistic} mechanics for the electron indicated that the
classical hydrogen atom would spontaneously ionize, use of
\textit{relativistic} mechanics for the electron might still lead to a stable
hydrogen ground state within classical electrodynamics. \ 

\subsection{Calculations of Nieuwenhuizen and Liska}

Recently, Nieuwenhuizen and Liska published their much more extensive
numerical calculations for the ground state of hydrogen within classical
electrodynamics including classical zero-point radiation.\cite{NL1}\cite{NL2}
\ Their numerical simulations involved full three-dimensional motion for the
nonrelativistic electron and are extended to far longer times than those
reported by Cole and Zou. \ According to the simulations, the balance between
radiation energy loss and energy gain from the zero-point radiation gives
millions of orbits for the electron in the approximate neighborhood of the
Bohr orbit without any indication that the electron falls into the potential
center. \ However, Nieuwenhuizen and Liska report that ionization of the
electron always occurs in their simulations.\cite{ionization} \ Furthermore
the distributions of radial position and energy are only in very rough
agreement with the Schroedinger ground-state results. \ Once again, the
ionization is reported as arising from the plunging orbits of low angular
momentum. \ For these plunging orbits, the multiply periodic orbit expansions
required high harmonics above the fundamental frequency. \ \ For these
plunging orbits, both the energy loss at the higher harmonics will be large,
and also the zero-point radiation energy gain from the higher harmonics will
be large because the zero-point spectrum increases with frequency. \ Indeed,
the figures of Nieuwenhuizen and Liska show spikes in energy which seem to
occur at the same time as spikes to large values of orbital eccentricity,
$\epsilon\rightarrow1$, consistent with the idea of large energy changes for
plunging orbits.\cite{figs}

Apparently Nieuwenhuizen and Liska were aware of the proposal\cite{B-h} that
use of relativistic trajectories for the electron would smooth the plunging
trajectories and so modify the energy pick up and loss associated with the
higher harmonics. \ In their second publication,\cite{NL2} they introduced the
lowest-order relativistic correction to the nonrelativistic energy and carried
out a new calculation using this corrected expression for the electron motion,
where they also included terms involving electron spin. \ Their report is that
these lowest-order relativistic corrections made little difference in their
calculations. \ The ionization of the electron was still observed.

The question as to whether or not the presence of classical zero-point
radiation leads to a stable ground state for hydrogen seems important. \ In
order to understand nature, physicist need to know the areas of agreement and
disagreement between classical and quantum theories. \ The calculations of
Nieuwenhuizen and Liska present an important improvement in understanding the
classical description of the hydrogen ground state. \ Nevertheless, given the
approximations in their calculations, one may wonder whether their conclusions
regarding the ionization are indeed justified.

\section{Role of Relativity in Modern Physics}

\subsection{When is Relativity Needed?}

Contemporary physics regards all of nature as relativistic in its fundamental
interactions. \ Indeed, electromagnetism is a relativistic theory. \ However,
most physicists believe that relativistic physics is needed for mechanical
systems only when dealing with particles whose speeds approach the speed of
light relative to the laboratory. \ Thus for atomic electron speeds in
hydrogen, we have $v\approx e^{2}/\hbar\approx c/137,$ so that nonrelativistic
physics is usually deemed adequate, and the electron orbits in classical
physics should correspond to the familiar conic sections: ellipse, parabola,
and hyperbola. \ However, this confidence in the adequacy of nonrelativistic
calculations is sometimes misplaced. \ 

\subsection{Relativistic Mechanical Kepler Orbits}

Thus it comes as a shock to many physicists to learn that the plunging Kepler
orbits of small angular momentum are radically different in relativistic
versus nonrelativistic physics. \ This difference holds even for particles
which have initial trajectories of arbitrarily speed, including very small
speed. \ Indeed, using relativistic mechanical orbits and ignoring any
radiation energy loss, if the orbital angular momentum $J$ is less than
$Ze^{2}/c$ (and irrespective of the particle energy), then the particle will
plunge into the Coulomb center while conserving energy and angular
momentum.\cite{B-orbs} \ This behavior is totally different from the situation
in nonrelativistic mechanics. \ In nonrelativistic physics, mechanical Kepler
orbits never plunge into the potential center unless the angular momentum is
exactly zero.

A sense of the change in perspective between relativistic and nonrelativistic
mechanical orbits is given easily by considering a circular Kepler orbit at
radius $r$ for a particle of mass $m$. \ In the relativistic case, we have
Newton's second law for a circular orbit
\begin{equation}
m\gamma\frac{v^{2}}{r}=\frac{Ze^{2}}{r^{2}}%
\end{equation}
where the angular momentum $J$ is given by
\begin{equation}
J=rm\gamma v.
\end{equation}
\ Therefore, combing these two equations, we have
\begin{equation}
Jv=Ze^{2}\text{ \ or \ }v=Ze^{2}/J.
\end{equation}
$~$\ This last equation holds also in the nonrelativistic case which we can
calculate by omitting the relativistic factor of $\gamma=(1-v^{2}%
/c^{2})^{-1/2}.$ \ However, for the relativistic orbit, the speed $v$ is
limited by $c$, so that $v=Ze^{2}/J<c.$ \ But then for any circular orbit, we
must have
\begin{equation}
Ze^{2}/c<J.
\end{equation}
\ This limit does not exist in nonrelativistic classical physics where
circular Kepler orbits of arbitrarily small positive angular momentum are
possible. \ As we have mentioned above, if $J\leq Ze^{2}/c,$ then no
relativistic circular orbit exists, and the particle will plunge into the
potential center while conserving energy and angular momentum. \ Indeed, the
critical angular momentum $Ze^{2}/c$ appears in a problem in Goldstein's
Classical Mechanics.\cite{Goldstein} \ The relativistic mechanical energy $E$
for a relativistic particle of mass $m$ in a Coulomb potential $Ze^{2}/r$ when
expressed in terms of action-angle variables $J_{2}$ and $J_{3}$ (which
include angular momentum) is given as\cite{2pi}
\begin{equation}
\frac{E}{mc^{2}}=\left(  1+\left[  \frac{J_{3}c}{Ze^{2}}-\frac{J_{2}c}{Ze^{2}%
}+\left\{  \left(  \frac{J_{2}c}{Ze^{2}}\right)  ^{2}-1\right\}
^{1/2}\right]  ^{-2}\right)  ^{-1/2}%
\end{equation}
\ We notice that if $J_{2}<Ze^{2}/c,$ then the energy expression involves the
square-root of \ a negative quantity; this is the signal that the periodic
behavior assumed in defining the action-angle variable $J_{2}$ no longer
holds, because the trajectories plunge into the Coulomb center.\cite{QM}

\subsection{Blackbody Radiation}

Within classical physics, the problem of atomic collapse is intimately bound
up with the problem of the blackbody radiation spectrum. \ Thus the
equilibrium spectrum of random classical radiation (blackbody radiation) must
be stable under scattering by a classical scattering system. \ The hydrogen
atom should provide an example of a scattering system in nature. \ Thus the
hydrogen atom should scatter the radiation, and radiation should in turn
provide the structure of the hydrogen atom. \ From this perspective,
zero-point radiation, which is the ground state of the random radiation,
should be in equilibrium with the scattering provided by the ground state of
hydrogen. \ 

The classical blackbody spectrum provides an example of the crucial importance
of special relativity. \ Understanding the Planck blackbody spectrum within
classical physics requires the use of a relativistic analysis; use of
nonrelativistic theory gives only the Rayleigh-Jeans low-frequency limit.
\ For example, it is a familiar observation in modern physics classes that the
use of the equipartition theorem from nonrelativistic statistical mechanics
leads to the Rayleigh-Jeans spectrum.\cite{RJ} \ Furthermore, use of
nonrelativistic scatterers to obtain the equilibrium spectrum of random
classical radiation also leads to the Rayleigh-Jeans spectrum.\cite{B-scat}

Since nonrelativistic scatterers give only the Rayleigh-Jeans spectrum as the
stable spectrum under scattering, one wishes to consider relativistic
scatterers. \ However, many physicists are unaware that relativistic physics
places very strong restrictions on the allowed mechanical systems. \ Most
mechanical systems have no relativistic extension consistent with
electromagnetism. \ The \textquotedblleft no-interaction
theorem\textquotedblright\ of Currie, Jordan, and Sudarshan\cite{CJS} requires
that any mechanical interaction beyond point collisions requires a field
theory. \ It is precisely on account of this restriction that elementary
classroom discussions of relativistic particle interactions always involve
point collisions and never interactions through a potential.\cite{B-direct}
\ The Coulomb potential is the only classical mechanical potential which has
been extended to a relativistic field theory, namely classical
electrodynamics. \ The scattering of random electromagnetic radiation in the
presence of a relativistic hydrogen atom indeed corresponds to a relativistic
field-theory situation which has all the qualitative aspects allowing
radiation equilibrium at the Planck spectrum.\cite{B-bb} \ Thus the mass $m$
of the charged particle in orbit about the Coulomb center provides the only
scale factor for length or frequency, while the action-angle variables are
invariant under adiabatic transformation of the strength $Ze^{2}$ of the
Coulomb potential.\cite{Adiabatic} \ Thus any situation of classical radiation
equilibrium in zero-point radiation for one mass $m$ and scale $Ze^{2}$ will
also hold for other masses $m^{\prime}$ or other choices of the constant
$Z^{\prime},$ while the action-angle variables will retain their values,
provided that Z is not too large. \ 

Furthermore, the full Planck spectrum can be obtained explicitly by use of
scaling symmetry within a relativistic setting. \ The zero-point radiation
spectrum is derived as the Lorentz-invariant spectrum in Minkowski
spacetime.\cite{M1965}\cite{B-rel} \ This spectrum is invariant under the
scale transformation appropriate for electromagnetic theory.\cite{Scale}
\ Thus electromagnetism is invariant under the scale transformation which
simultaneously multiplies all lengths and times by a positive constant
$\sigma$ (thus preserving $c=length/time)$ and all energies by $1/\sigma$
(thus preserving $e^{2}=energy\times length)$. \ Under this scale
transformation, thermal radiation at the temperature $T$ is transformed to
thermal radiation at temperature $T/\sigma;$ but zero-temperature zero-point
radiation is unchanged in Minkowski spacetime, consistent with zero
temperature divided by any positive constant $\sigma$ still being zero
temperature. \ However, zero-point radiation, which is scale invariant in
Minkowski spacetime, acquires a local scale in an accelerated Rindler
coordinate frame.\cite{Rindler} \ When a scale transformation is applied to
zero-point radiation in a Rindler frame, the zero-point radiation spectrum is
carried into thermal radiation at non-zero temperature. \ \ If we now imagine
moving ever further away from the Rindler event horizon while applying a
rescaling so as to keep our local temperature constant, then we move to a
region where the acceleration vanishes, and we recover exactly the Planck
spectrum at finite temperature in Minkowski spacetime.\cite{Rindler} \ We note
that this analysis requires a fully relativistic theory at every
step.\cite{Also}

\section{Conclusion}

Today our classes in modern physics still hear mention of the problem of
atomic collapse as an example of the failure of classical theory to account
for atomic structure. \ Furthermore, some quantum descriptions state that
zero-point motion prevents the collapse. \ However, most classes receive no
mention of the possibility of classical electromagnetic zero-point radiation
causing the zero-point motion, despite the fact that classical zero-point
radiation gives at least a heuristic classical idea regarding many phenomena
and an exact classical account of some phenomena which are currently regarded
as \textquotedblleft quantum phenomena.\textquotedblright\ \ Thus Casimir
forces, van de Waals forces, oscillator behavior, oscillator specific heats,
blackbody radiation, and diamagnetism all have unimpeachable classical
calculations which give results in exact agreement with the corresponding
quantum results.

Regarding the problem of atomic collapse, we are currently in an interesting
new regime. \ The classical problem of atomic collapse mentioned in our modern
physics classes assumes that there is no random classical radiation present to
give energy to the radiating electron which is losing energy. \ When classical
electromagnetic zero-point radiation (which gives us results for Casimir
forces identical to those obtained from quantum electrodynamics) is applied to
the classical hydrogen atom, the traditional problem of atomic collapse
disappears, and the electron indeed orbits the Coulomb center near the Bohr
radius for millions of orbits without falling into the potential center.
\ However, the new numerical calculations suggest that the classical problem
of atomic collapse has been replaced by the problem of the zero-point
radiation delivering too much energy to the orbiting electron so as to cause
self-ionization of the atom. \ But this self-ionization problem may be linked
to the question of using a nonrelativistic analysis. \ An analytic calculation
for the relativistic hydrogen atom in zero-point radiation has never been
done. \ At the present time, the numerical simulations for the classical
hydrogen atom are exceedingly difficult, even though they are not fully
relativistic. \ We still do not know whether or not classical zero-point
radiation provides an acceptable ground state or causes self-ionization of the
classical hydrogen atom.\


\begin{thebibliography}{99}
\bibitem {texts}\ In the textbooks, classical electromagnetism is treated from
the point of view of electromagnetic technology where the charge and current
sources provide overwhelmingly large electromagnetic fields; classical
electromagnetism is not treated as a subject relevant for atomic physics where
the sources may give fields comparable to the background fields of thermal
radiation or zero-point radiation. \ Accordingly, the homogeneous boundary
conditions on Maxwell's equations are rarely if ever mentioned in textbooks of
classical electromagnetism.
\bibitem {ionization}Nieuwenhuizen and Liska define ionization
\textquotedblleft as the moment when the electron stays above $\mathcal{E}%
=-0.05$ for a duration for at least 10$^{7}$t$_{0}$.\textquotedblright\ \ Here
$\mathcal{E}$ is given in Bohr units and t$_{0}$ is the Bohr period.
\bibitem {2pi}Here we have chosen action-angle variables differing by a factor
of $1/(2\pi)$ from those of problem 28 of Goldstein's text and have written
the strength of the potential as $k=Ze^{2}.$
\bibitem {Adiabatic}This adiabatic invariance corresponds to the familiar
mechanical problem of slowly pulling a string through a hole in a frictionless
table, with the string attached to a rotating block. \ The angular momentum
$J$ (an action variable) is conserved during the slow change of the strength
of the central force.
\end{thebibliography}
\end{document}